\documentclass[11pt,a4paper]{article}
\pdfoutput=1
\usepackage{jcappub}
\usepackage{amsfonts}
\usepackage{amssymb}
\usepackage{epsfig}
\usepackage{ifpdf}
\usepackage{graphicx}
\usepackage{braket}
\usepackage{amsmath}

\title{Axion-like particle oscillations}
\author[1]{Francesca Chadha-Day}
\affiliation[1]{Department of Physics, Durham University, South Road, Durham DH1 3LE, United Kingdom}
\abstract{

String theory compactifications may generate many light axion-like particles (ALPs) with weak couplings to electromagnetism. In general, a large number of ALPs may exist, with a linear combination having a potentially observable coupling to electromagnetism. The basis in which only one ALP couples to electromagnetism is in general misaligned with the mass basis. This leads to mixing between the `electromagnetic' ALP and a number of `hidden' ALPs that do not interact directly with the photon. The process is analagous to neutrino oscillations. I will discuss the phenomenological consequences of this mixing for astrophysical ALP signals, in particular showing that it may significantly reduce the predicted signal in experiments such as the CERN Axion Solar Telescope.
\hfill \\

}

\begin{document}
 \maketitle

\section{Introduction}

The existence of the QCD axion is well motivated by the strong CP problem \cite{Peccei:1977hh,Wilczek:1977pj}. The QCD axion also arises in string theory models, along with a potentially large number of axion-like particles (ALPs) that do not couple directly to gluons \cite{hep-th/0602233, hep-th/0605206, 1206.0819,2105.02889}. Significant experimental and theoretical effort has been dedicated to detecting these ALPs via their coupling to photons, either through astrophysical observations or through ground based experiments \cite{ParticleDataGroup:2020ssz}. In particular, in the presence of a background magnetic field, ALPs and photons may interconvert, leading to striking and potentially detectable signatures \cite{Raffelt:1987im}. 

Recent work has also explored the distinctive phenomenology of the string {\it axiverse} \cite{0905.4720} in astrophysical and cosmological contexts - many ALPs with different masses. For example, string axiverse phenomenology has been studied in the context of black hole superradiance \cite{1805.02016,2103.06812} and the cosmology of ALP dark matter \cite{1706.03236,2104.09923}. The phenomenology of multiple axions or ALPs has also been studied in the context of inflation \cite{hep-ph/0409138,hep-th/0507205}, kinetic mixing between the QCD axion and ALPs \cite{1709.06085}, isocurvature perturbations \cite{1411.2011}, cosmic birefringence \cite{2108.02150,2103.06812} and decay of axion dark matter \cite{1408.3936}.

In this work, I will consider astrophysical string axiverse phenemenology in the context of electromagnetic detection of ALPs. As shown in \cite{1909.05257}, we can move to a basis in which a linear combination of ALPs couple to electromagnetism. The remaining ALPs are `hidden' with no direct coupling to photons. I will show that this basis transformation leads to a mass mixing between the electromagnetic ALP and the hidden ALPs. This mixing may have a significant effect on a subset of ALP detection experiments.

This paper is structured as follows. In section 2, I will introduce the Lagrangian for ALPs in the string axiverse. In section 3, I will consider moving to the electromagnetic basis, in which only one ALP couples directly to photons, and discuss ALP oscillations in this basis.  In section 4, I will discuss the observational consequences of these oscillations. In section 5, I will conclude and discuss the implications of this work.

\section{The String Axiverse}

We will consider a string axiverse scenario with $N$ axion fields $\phi^i$. Each $\phi^i$ would typically interact rather weakly with electromagnetism, but the combined effect of many such fields leads to a potentially observable coupling between the axion sector and the photon \cite{1909.05257}. The $\phi^i$ also in general couple to gluons as $\mathcal{L} \supset g_{i}^{g} \phi^{i} G_{\mu \nu} \tilde{G}^{\mu \nu}$. We therefore have:

\begin{equation}
\begin{split}
\mathcal{L} \supset \sum_i \Bigl( & -\frac{1}{2} \partial^{\mu} \phi^{i} \partial_{\mu} \phi^i - \frac{1}{2} m_i^2 (\phi^i)^2  \\
&  -g^{\gamma}_i \phi^i \tilde{F}^{\mu \nu} F_{\mu \nu} -g^{g}_i \phi^i \tilde{G}^{\mu \nu} G_{\mu \nu} \Bigl)
\end{split}
\end{equation}

We will move to a basis in which only one field, $\phi_{\rm QCD}$, couples to gluons. This field is then the QCD axion \cite{1206.0819}. The mass and coupling to photons of $\phi_{\rm QCD}$ are set by its interaction with the QCD sector, in particular by mixing with the neutral pion \cite{1511.02867}. For the remaining fields, which do not couple to directly to gluons, their mass and coupling to photons are in principle independent parameters. We will therefore refer to these as axion-like particles or ALPs. Note that we are no longer in the mass basis, and so we expect some mixing between  $\phi_{\rm QCD}$ and the other $\phi^i$. However, as the mass of the QCD axion is generated by QCD effects, we will assume that the QCD basis is approximately aligned with the mass basis, and therefore mass mixing between the QCD axion and the ALPs is negligible. We are free to choose a basis where the mass matrix for the ALPs is diagonal. We therefore have the ALP Lagrangian 

\begin{equation}
\label{bigEquation}
\mathcal{L} \supset \sum_i \left( -\frac{1}{2} \partial^{\mu} \phi^{i} \partial_{\mu} \phi^{i} -  \frac{1}{2} m_i^2 (\phi^i)^2-g^{\gamma}_i \phi^i \tilde{F}^{\mu \nu} F_{\mu \nu} \right),
\end{equation}

where the $\phi_i$ have been redefined such that they are orthogonal to $\phi_{\rm QCD}$. In this paper, we will discuss the electromagnetic phenomenology of the Lagrangian in equation \ref{bigEquation}. The parameters $m_i$ and $g^{\gamma}_i$, as well as the total number of ALPs depend on the details of the string compactification. The purpose of this paper is therefore to draw attention to the high level of model dependence in interpreting and comparing different bounds on ALPs.

\section{ALP oscillations in the electromagnetic basis}

As described in \cite{1909.05257}, we will now move to the electromagnetic basis in which only one ALP couples directly to $\tilde{F}^{\mu \nu} F_{\mu \nu}$. Let this electromagnetic ALP be $\phi^1$. We then have

\begin{equation}
\phi^1 = \frac{\sum_i g^{\gamma}_i \phi^i}{\sqrt{\sum_i {g^{\gamma}_i}^2}}.
\label{transformation}
\end{equation}

We redefine the other ALP fields such that they are orthogonal to $\phi^1$. In this electromagnetic basis, the mass matrix is not diagonal. The Lagrangian is

\begin{equation}
\mathcal{L} \supset - \sum_i \frac{1}{2} \partial^{\mu} \phi^{i} \partial_{\mu} \phi^{i} -  \sum_{i,j} \frac{1}{2} M_{i j} \phi^i \phi^j -g^{\gamma} \phi^1 \tilde{F}^{\mu \nu} F_{\mu \nu} ,
\end{equation}

where $g^{\gamma} =\sqrt{\sum_i (g^{\gamma}_i)^2}$. As shown in \cite{1909.05257}, $g^{\gamma}$ may be large enough to be observable. Note that $\phi_1$ now has a mass mixing with each of the other hidden ALPs. We may, however, choose a basis such that there is no mass mixing within the hidden ALP sector.
 
Moving to the electromagnetic basis leads to oscillations between the electromagnetic ALP and the hidden ALPs. The effect is analagous to neutrino oscillations. This effect occurs when an ALP is produced via an electromagnetic process such as the Primakoff effect. In the case where ALP masses are negligible, as considered in \cite{1909.05257}, the produced ALP will be $\phi^1$ - the linear combination of ALP fields that couple to electromagnetism. This is analagous to the fact that weak processes produce neutrinos in flavour eigenstates. However, $\phi^1$ is not a mass eigenstate, and therefore will oscillate into the hidden ALPs as it propagates.

The transformation between the mass and electromagnetic bases is given by

\begin{equation}
\ket{\phi^i_{\rm mass}} = U_{j i} \ket{\phi^j_{\rm EM}},
\end{equation}

where $U$ is a unitary matrix that includes the transformation in equation \eqref{transformation}. In the ultra-relativistic limit, the mass eigenstates $\phi^i_{\rm mass}$ propagate as $\ket{\phi^i_{\rm mass}(L)} = {\rm e}^{-i \frac{m_i^2 L}{2 E}} \ket{\phi^i_{\rm mass}(0)}$. Following the standard derivation for neutrino oscillations, if the probability of interconverting with the photon is negligible, the survival probability of the electromagnetic ALP $\ket{\phi^1_{\rm EM}}$ after propagating a distance $L$ in vacuum is then

\begin{equation}
P_{1 \rightarrow 1} = |\braket{\phi^1_{\rm EM}(L) | \phi^1_{\rm EM}(0)}|^2 = \left| \sum_i U^{\star}_{1 i} U_{1 i}  {\rm e}^{-i \frac{m_i^2 L}{2 E}} \right|^2.
\end{equation}

This gives

\begin{equation}
\label{suvivalGeneral}
P_{1 \rightarrow 1} = 1 - 4 \sum_{i > j} |U_{1 i}|^2 |U_{1 j}|^2 {\rm sin}^2 \left( \frac{\Delta m_{ij}^2 L}{4 E} \right),
\end{equation}
where $\Delta m_{ij}^2 \equiv m_i^2 - m_j^2$.
 
 \section{The phenomenology of ALP oscillations}
 
Calculations of the phenomenology of the ALP-photon interaction have in general considered only one ALP coupled to photons, and have not considered the mass mixing between this electromagnetic ALP and any hidden ALPs. However, in some experiments this mass mixing may significantly affect the predicted signal. In particular, this effect is relevant in situations where an ALP is produced via an electromagnetic interaction and then propagates a large distance before being detected electromagnetically. In this case, we must take into account the probability that the electromagnetic ALP will oscillate into an undetectable hidden ALP. 

We will calculate the effect of mixing with hidden ALPs on the predicted ALP signal from the Cern Axion Solar Telescope (CAST) \cite{1705.02290} and from the supernova SN1987A \cite{1410.3747}. In both cases, we will assume that the QCD axion is too heavy to lead to a significant signal, such that only the ALPs are relevant. In models with a light QCD axion and an ALP coupled to electromagnetism, this would lead to an additional contribution to the signal. CAST seeks to detect axions and ALPs produced in the Sun via their interconversion with photons in a laboratory magnetic field. In the case of ALP production in the Sun via the ALP coupling to photons, the electromagnetic ALP produced may oscillate into a hidden ALP as it propagates towards Earth. As these hidden ALPs would produce no signal in CAST, we expect this effect to lead to a reduction in the predicted CAST signal for a given $g^{\gamma}$ when compared to the case with only a single ALP field with coupling $g^{\gamma}$ to photons and no hidden ALPs.

The supernova SN1987A was observed in the nearby Large Magellanic Cloud in 1987. Low mass ALPs would have been copiously produced in the supernova, primarily via their electromagnetic interaction. The produced ALPs would therefore again be the linear combination of light ALP fields that couple to photons. These ALPs would interconvert with photons in the Milky Way magnetic field, leading to the production of observable gamma rays. The non-observation of these gammas rays is used in  \cite{1410.3747} to place bounds on $g^{\gamma}$. However, if the electromagnetic ALP oscillates into a hidden ALP between the Large Magellanic cloud and the Milky Way, this hidden ALP would not mix with the photon in the Milky Way magnetic field, and hence the expected gamma ray signal for a given $g^{\gamma}$ would be reduced.

Other observations would be much less sensitive to ALP oscillations into hidden ALPs. For example, ALP induced stellar cooling \cite{1406.6053} is sensitive primarily to the production rate of ALPs, and will not be significantly affected if these ALPs subsequently oscillate into hidden ALPs. Similarly, ALP induced modulations in point source spectra \cite{1907.05475,1704.05256,1603.06978} in the low conversion probability regime result primarily from photon to ALP conversion, and hence the effect of subsequent oscillations into hidden ALPs will be small.

I will now turn to consider the expected ALP signal from CAST and from SN1987A in the case of a single hidden ALP and in the case of a large number of hidden ALPs.
 
 \subsection{The two ALP case}
 We will first consider the case where there are only two ALPs with non-negligible coupling to photons. The Lagrangian in the mass basis is

\begin{equation}
\begin{split}
\mathcal{L} \supset & -\frac{1}{2}  \partial^{\mu} \phi_{1} \partial_{\mu} \phi_{1} -\frac{1}{2}  \partial^{\mu} \phi_{2} \partial_{\mu} \phi_{2}  \\
& -\frac{1}{2} m_1^2 \phi_1^2  -  \frac{1}{2} m_2^2 \phi_2^2-g^{\gamma}_1 \phi_1 \tilde{F}^{\mu \nu} F_{\mu \nu}  -g^{\gamma}_2 \phi_2 \tilde{F}^{\mu \nu} F_{\mu \nu}.
\end{split}
\end{equation}

We will now move to the basis where only one ALP couples to photons. The electromagnetic ALP is

\begin{equation}
\phi_{\gamma} = \frac{g^{\gamma}_1 \phi_1 + g^{\gamma}_2 \phi_2}{\sqrt{{g^{\gamma}_1}^2 + {g^{\gamma}_2}^2}}.
\end{equation}

The hidden ALP is

\begin{equation}
\phi_{h} = \frac{g^{\gamma}_2 \phi_1 - g^{\gamma}_1 \phi_2}{\sqrt{{g^{\gamma}_1}^2 + {g^{\gamma}_2}^2}}.
\end{equation}

The mass basis and the electromagnetic basis are related by

\begin{equation}
\begin{pmatrix}
           \phi_{1} \\
           \phi_{2} \\
         \end{pmatrix} = \begin{pmatrix} {\rm cos} \theta & - {\rm sin} \theta \\  {\rm sin} \theta & {\rm cos} \theta \end{pmatrix} \begin{pmatrix} \phi_{\gamma} \\ \phi_h \end{pmatrix},
 \end{equation}
 
with

\begin{equation}
\theta = {\rm cos}^{-1} \left( \frac{g_1^{\gamma}}{\sqrt{{g_1^{\gamma}}^2 + {g_2^{\gamma}}^2}} \right).
\end{equation}

In the absence of a background magnetic field (and therefore mixing with the photon), the probability of an electromagnetic ALP oscillating into a hidden ALP is

\begin{equation}
P_{\phi_{\gamma} \rightarrow \phi_{h}} = {\rm sin}^2{2 \theta} {\rm sin}^2\left(\frac{\Delta m^2 L}{4 E}\right),
\end{equation}

where $\Delta m^2 = m_1^2 - m_2^2$, $L$ is the propagation length and $E$ is the ALP's energy. This oscillation probability is exactly analogous to that for neutrino oscillations with two flavours.

\subsubsection{CAST: The two ALP case}

The Cern Axion Solar Telescope (CAST) \cite{1705.02290} seeks to observe axions and ALPs produced in the Sun by inducing their conversion to photons in a magnetic field on Earth. We will assume that $m_1$ and $m_2$ are both negligible compared to the temperature and effective photon mass in the Sun. In this case, Primakoff production in the Sun produces the state $\phi_{\gamma}$. There will be no significant Primakoff production of the hidden ALP as it has no direct coupling to photons. As described in \cite{1909.05257}, the production of the electromagnetic ALP in the Sun in this case will not be significantly altered from the case of a single mass eigenstate ALP with negligible mass and with coupling to photons $g^{\gamma} = \sqrt{{g_1^{\gamma}}^2+{g_2^{\gamma}}^2}$. As described above, ALPs produced electromagnetically in the Sun may oscillate into hidden ALPs as they travel to Earth, and therefore be unobservable to CAST. For a single hidden ALP, the electromagnetic ALP survival probability is given by

\begin{equation}
P_{\phi_{\gamma} \rightarrow \phi_{\gamma}} = 1 -{\rm sin}^2 2 \theta \frac{\int_{2 \, {\rm keV}}^{7 \, {\rm keV}} {\rm sin}^2 \left( \frac{\Delta m^2 L}{4 E} \right) \frac{d \Phi_a}{d E} dE}{\int_{2 \, {\rm keV}}^{7 \, {\rm keV}} \frac{d \Phi_a}{d E} dE},
\end{equation}
where $L$ is the Earth-Sun distance and $\frac{d \Phi_a}{d E}$ is the differential solar ALP flux as given in  \cite{1705.02290}. We integrate this flux over the CAST energy range of interest. The expected CAST signal will be reduced in proportion to $P_{\phi_{\gamma} \rightarrow \phi_{\gamma}}$. Note we have neglected the extremely small probability of ALP to photon conversion during propagation from the Sun to the Earth \cite{0804.3543}.

\begin{figure} [h]
\includegraphics[scale=1]{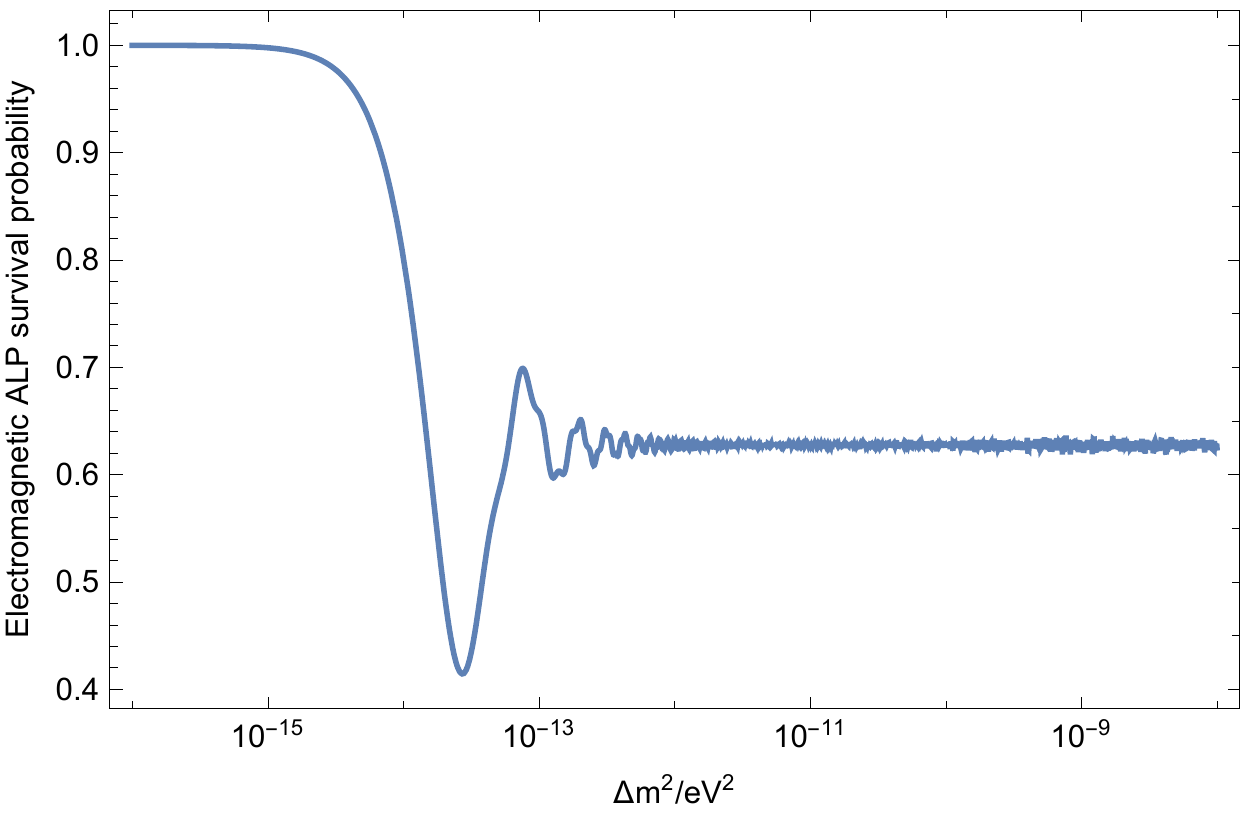}
\caption{The survival probability for electromagnetic ALPs observed by CAST for a mixing angle $\theta = \frac{\pi}{3}$.}
\label{CastPlot}
\end{figure}

The resulting electromagnetic ALP survival probability is shown in Figure \ref{CastPlot} for a mixing angle $\theta = \frac{\pi}{3}$ and a range of mass differences. We see that the expected CAST signal is significantly reduced for $\Delta m^2 \gtrsim 10^{-14} \, {\rm eV}^2$. For $\Delta m^2 \gtrsim 10^{-12} \, {\rm eV}^2$, oscillations in the survival probability are washed out by averaging over the ALP flux, and we have $P_{\phi_{\gamma} \rightarrow \phi_{\gamma}} \simeq 1 - \frac{1}{2}{\rm sin}^2 2 \theta$.

\subsubsection{SN1987A: The two ALP case}
A similar effect may be seen in observations of ALPs from SN1987A, the 1987 supernova in the nearby Large Magellanic Cloud. ALPs would have been produced during the supernova via their interaction with the photon, and subsequently converted to gamma-rays in the Milky Way magnetic field. The non-observation of these gamma-rays has been used to place bounds on the ALP-photon coupling \cite{1410.3747}.

\begin{figure} [h]
\includegraphics[scale=1]{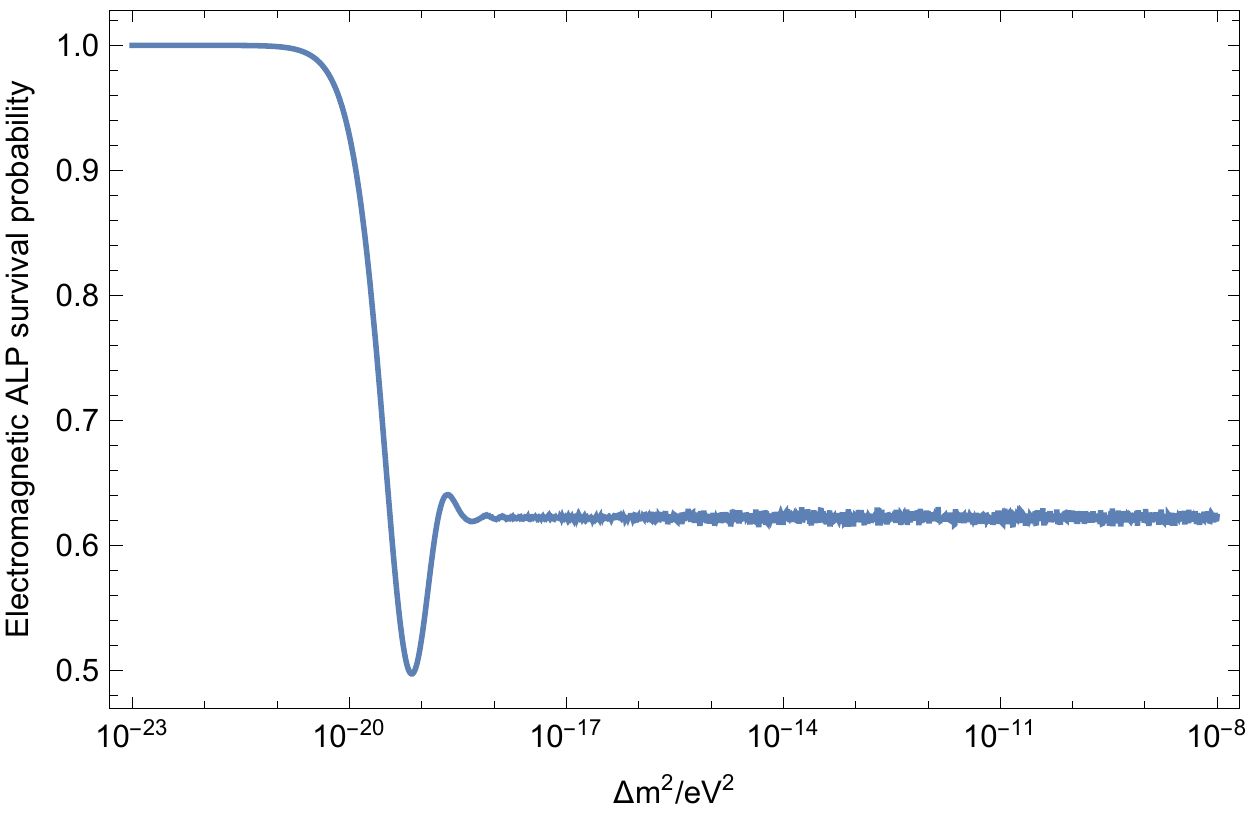}
\caption{The survival probability for electromagnetic ALPs travelling from the Large Magellanic Cloud to the Milky Way for a mixing angle $\theta = \frac{\pi}{3}$.}
\label{SNPlot}
\end{figure}

If the electromagnetic ALPs produced in the supernova oscillated into hidden ALPs between the Large Magellanic Cloud and the Milky Way, the gamma-ray signal would be correspondingly reduced. The electromagnetic ALP survival probability in the case of a single hidden ALP is shown in Figure \ref{SNPlot}. We again average over the differential axion flux as given in \cite{1410.3747}, which has typical energies $\mathcal{O}(100 \, {\rm MeV})$. In this case, we have $P_{\phi_{\gamma} \rightarrow \phi_{\gamma}} \simeq 1 - \frac{1}{2}{\rm sin}^2 2 \theta$ for $\Delta m^2 \gtrsim 10^{-18} \, {\rm eV}^2$.

 \subsection{The many ALP case}
Let us now consider a situation where many ALP mass eigenstates have non-negligible coupling to electromagnetism, leading to many hidden ALPs. We will remain agnostic as to the specifics of the ALP mass spectrum. The electromagnetic ALP survival probability is then given by equation \eqref{suvivalGeneral}. In the long $L$ limit for any given $\Delta m_{ij}^2$, $L \gg \frac{4 E_0}{\Delta m_{ij}^2}$, we obtain
 
 \begin{equation}
 \frac{\int{\rm sin}^2 \left( \frac{\Delta m_{ij}^2 L}{4 E} \right) \frac{d \Phi_a}{d E} dE} {\int \frac{d \Phi_a}{d E} dE} \sim \frac{1}{2},
 \end{equation}
 where $\frac{d \Phi_a}{d E}$ is the differential ALP flux of the source and $E_0$ is the average ALP energy. In the long $L$ limit for {\it all} $\Delta m_{ij}^2$, we therefore have an electromagnetic ALP survival probability
 
 \begin{equation}
 P_{1 \rightarrow 1} \sim 1 - 2 \sum_{i>j} |U_{1i} U_{1j}|^2.
 \end{equation}
 
 Now, from the definition of the electromagnetic ALP we have
 
 \begin{equation}
 U_{1 i} = \frac{g^{\gamma}_i}{\sqrt{\sum_{i} {g^{\gamma}_i}^2}},
 \end{equation}
 
 giving
 
  \begin{equation}
  \begin{split}
 P_{1 \rightarrow 1} & \sim 1 - 2 \sum_{i>j} \left| \frac{g^{\gamma}_i g^{\gamma}_j}{\sum_{k} {g^{\gamma}_k}^2} \right|^2 \\
 & = 1 - 2 \frac{\sum_{i>j} \left( g^{\gamma}_i g^{\gamma}_j \right)^2}{\sum_{i>j} \left( g^{\gamma}_i g^{\gamma}_j \right)^2 + \sum_{j>i} \left( g^{\gamma}_i g^{\gamma}_j \right)^2 + \sum_{i=j} \left( g^{\gamma}_i g^{\gamma}_j \right)^2},
 \end{split}
 \end{equation}
 
assuming the $g_i^{\gamma}$ are real. For large numbers of hidden ALPs, the third term in the denominator is negligible, and so we have $P_{1 \rightarrow 1} \rightarrow 0$ for $L \gg \frac{4 E_0}{\Delta m_{ij}^2}$. For CAST observations, this corresponds to $\Delta m_{ij}^2 \gg 10^{-14} \, {\rm eV}^2$ for a large number of ALP mass eigenstates that contribute to the electromagnetic ALP. For observations of SN1987A, this corresponds to $\Delta m_{ij}^2 \gg 5 \times 10^{-20} \, {\rm eV}^2$ for a large number of ALP mass eigenstates that contribute to the electromagnetic ALP.

 \section{Discussion}
 
 In string axiverse scenarios, the axion-like particle that mixes with the photon may not be a mass eigenstate but a linear combination of many mass eigenstate ALPs \cite{1909.05257}. We have shown that this would lead to mass mixing between the electromagnetically coupled ALP and other `hidden' ALPs that do not couple directly to electromagnetism. Depending on the parameters of the ALP fields, we have shown that the observational effects are potentially dramatic. Note that we have not included any additional interaction between the ALP fields - the effect is purely due to basis misalignment. 
 
 We expect the most significant effects in settings where an ALP is produced electromagnetically and then {\it travels a large distance} before being (hopefully) detected electromagnetically. In these cases, mixing of the electromagnetic ALP with undetectable hidden ALPs during propagation may significantly reduce the expected signal. We have shown how this effect is applied to both helioscope and supernova bounds on the ALP-photon coupling. Where there is significant mixing between the electromagnetic and hidden ALPs, the bounds on $g^{\gamma}$ from these experiments could be substantially reduced. 
 
Further study of the effects of mixing with hidden ALPs on other search strategies such as stellar cooling and spectral modulations is required. However, there is no obvious mechanism for a dramatic reduction in signal strength in these cases. This means that oscillations with hidden ALPs could reconcile the possible observation of ALP induced modulations in galactic pulsar spectra with the bound on $g^{\gamma}$ from CAST \cite{1801.08813,2008.08100}. We also emphasise that this work does not apply to the QCD axion, whose mass and coupling to photons each derive in part from the coupling to QCD.

An analogous effect could occur with other ALP couplings, for example the ALP-fermion coupling. In fact, the mass, electromagnetic coupling and fermion coupling bases may all be mutually misaligned. This could have interesting phenomenological consequences, particularly for effects where both the electromagnetic and fermionic couplings are relevant, such as considered in \cite{1302.6283,1903.05088}.

We have shown that the phenomenology of a string axiverse may be quite different to that of a single ALP. We have shown here that mixing with hidden ALPs may lead to significant signal reductions in some experiments. More broadly, mass mixing in the string axiverse scenario means that bounds on $g^{\gamma}$ from different observations cannot be trivially compared. The interpretation of these bounds is highly model dependent. String axiverse phenomenology clearly warrants further study. 

\section*{Acknowledgements}
I am supported by Stephen Hawking Fellowship EP/T01668X/1 and STFC grant ST/P001246/1. I am very grateful to Sven Krippendorf for helpful comments on a draft of this paper. I further thank Rick Gupta, David J. E. Marsh, Jack Shergold and Jessica Turner for helpful discussions.
 
\bibliography{manyAxions}
\bibliographystyle{JHEP}

\end{document}